\documentclass[%
preprint,
reprint,
superscriptaddress,
 amsmath,amssymb,
aps,
pra,
floatfix,
]{revtex4-2}
\usepackage[dvipsnames]{xcolor}
\usepackage{graphicx}% Include figure files
\usepackage{dcolumn}% Align table columns on decimal point
\usepackage{bm}% bold math

\begin{document}

% \preprint{APS/123-QED}

\title{Optimal training of finitely-sampled quantum reservoir computers for forecasting of chaotic dynamics}

\author{Osama Ahmed}
\email{o.ahmed22@imperial.ac.uk}
\affiliation{Imperial College London, Department of Aeronautics, Exhibition Road, London SW7 2BX, UK} 

\author{Felix Tennie}
\affiliation{City St George's, University of London, Department of Engineering, Northampton Square
London EC1V 0HB, UK}
\affiliation{Imperial College London, Department of Aeronautics, Exhibition Road, London SW7 2BX, UK} 

\author{Luca Magri}%
\email{l.magri@imperial.ac.uk}

\affiliation{Imperial College London, Department of Aeronautics, Exhibition Road, London SW7 2BX, UK} 
\affiliation{The Alan Turing Institute, London NW1 2DB, UK}
\affiliation{Politecnico di Torino, DIMEAS, Corso Duca degli Abruzzi, 24 10129 Torino, Italy}

\begin{abstract}

In the current Noisy Intermediate Scale Quantum (NISQ) era, the presence of noise deteriorates the performance of quantum computing algorithms. Quantum Reservoir Computing (QRC) is a type of Quantum Machine Learning algorithm, which, however, can benefit from different types of tuned noise. In this paper, we analyse the effect that finite-sampling noise has on the chaotic time-series prediction capabilities of QRC and Recurrence-free Quantum Reservoir Computing (RF-QRC). First, we show that, even without a recurrent loop, RF-QRC contains temporal information about previous reservoir states using leaky integrated neurons. This makes RF-QRC  different from Quantum Extreme Learning Machines (QELM). Second, we show that finite sampling noise degrades the prediction capabilities of both QRC and RF-QRC while affecting QRC more due to the propagation of noise. Third, we optimize the training of the finite-sampled quantum reservoir computing framework using two methods: (a) Singular Value Decomposition (SVD) applied to the data matrix containing noisy reservoir activation states; and (b) data-filtering techniques to remove the high-frequencies from the noisy reservoir activation states. We show that denoising reservoir activation states improve the signal-to-noise ratios with smaller training loss. Finally, we demonstrate that the training and denoising of the noisy reservoir activation signals in RF-QRC are highly parallelizable on multiple Quantum Processing Units (QPUs) as compared to the QRC architecture with recurrent connections. The analyses are numerically showcased on prototypical chaotic dynamical systems with relevance to turbulence. This work opens opportunities for using quantum reservoir computing with finite samples for time-series forecasting on near-term quantum hardware.

\end{abstract}
\onecolumngrid

\maketitle

\section{\label{sec:intro}Introduction:\protect}

Despite various noise sources affecting the performance of quantum algorithms in NISQ devices, finite sampling noise is a device-independent  source of noise in various Quantum Machine Learning (QML) algorithms. Finite sampling noise, which provides a fundamental limit to learning in different QML applications \citep{hu2023tackling,mujal_time-series_2023}, is rooted in the foundations of quantum mechanics, and will be  present in  future Fault-tolerant Quantum computer (FTQC) as well \citep{preskill1998fault}.

Obtaining information on the density operator $\rho$ of a quantum system requires measurements of multiple copies of $\rho$, which is known as quantum state tomography \citep{9781107002173}. A complete quantum state tomography is exponentially hard and scales exponentially with the system size $\mathcal{O}(D^{2})$, where $ D = 2^{n} $ with $n$ qubits. In \citep{aaronson2018shadow}, they showed that instead of generating a full-classical description of quantum states, it is often sufficient to directly predict many properties of the associated quantum system efficiently (shadow tomography), for example in quantum chemistry and quantum simulations. The same analysis is extended by the approach of classical shadows \citep{huang2020predicting}, through which we can predict many associated properties of a quantum system from  few measurements. For QML applications, one often requires access to a large number of measurement expectation values to perform classification and regression tasks \citep{schuld2021machine}. Therefore, in those cases, the concept of shadow tomography does not result in useful QML models. 

The calculation of finite expectation values in variational quantum algorithms can also result in vanishing gradients and a mostly flat loss landscape called \textit{barren plateaus} \citep{mcclean2018barren}. To circumvent the issue of barren plateaus, Quantum Extreme Learning Machines (QELM) and Quantum Reservoir Computing (QRC) \citep{Fujii2021,pfeffer_hybrid_2022,mujal_opportunities_2021} are promising frameworks because they do not require the calculation of gradients for loss minimization. QRC is inspired by classical reservoir computers \citep{jaeger__2001} - a class of recurrent neural networks (RNNs), which have proven to be excellent tools for time series forecasting \citep{racca_data-driven_2022,10.1007/978-3-030-22747-0_15}. QELM, on the other hand, does not involve recurrence and is  easier to train but has limited applications. Most promising applications of QRC include forecasting chaotic dynamics \citep{ahmed2024prediction,fry2023optimizing} and quantum dynamics \citep{sornsaeng_quantum_2023} on a quantum computer.

QRC benefits from different types of tuned noises such as amplitude and phase damping noise \citep{domingo_taking_2023,fry_optimizing_2023}. QELM applications for time-series forecasting are, however, limited because they have no memory of past inputs and suffer from exponential concentration \citep{xiong2023fundamental}. A recent proposal of Recurrence-free QRC (RF-QRC) \citep{ahmed2024prediction} addresses both of these issues in a way that does not have recurrence built in the quantum circuit similar to QELM - making it easier to train. Instead, the information about previous reservoir states is fed as a classical post-processing step with leaky integrated neurons that have individual state dynamics and temporal memory \citep{JAEGER2007335}. Nevertheless, it may be possible to employ both QRC and RF-QRC in classical settings as a \textit{quantum-inspired} machine learning algorithm to improve the prediction capabilities of classical reservoir computers. For the prediction of chaotic dynamics and extreme events \citep{ahmed2024prediction}, this has been done by emulating the evolution of the quantum state vector, which allows us to numerically determine exact measurement expectation values. However, in order to realize any quantum advantage for the increasing number of qubits and for processing quantum data using QRC, we require the implementation of QRC (or RF-QRC) on quantum hardware with finite sampling.

Some previous proposals of QRC also consider the impact of finite sampling noise in physical implementations \citep{khan2021physical,dudas_quantum_2023}. A recent framework uses weak and projective measurements in QRC to reduce noise effects \citep{mujal_time-series_2023}. Another proposal for QML applications uses variance regularization to suppress probabilistic noise in the framework of quantum neural networks \citep{kreplin2024reduction}. Despite these proposals for suppressing finite-sampling noise, the effect of spreading of correlations for temporal learning tasks in recurrent QML applications still needs to be explored. 

In this work, we study and compare the impact of finite sampling noise on conventional QRC with recurrence and Recurrence-free QRC (RF-QRC) architectures. We focus our analysis on the finite sampling noise for two reasons: (i) 
The physical computing time required for executing quantum circuits imposes limitations on the possible number of shots taken for a learning task, which results in lower bounds on the size of finite sampling noise. (b) Beyond that, in some cases, QRC can instead benefit from certain types of tuned noises \citep{fry_optimizing_2023}. Therefore, the motivation for this work is to analyze the impact of sampling noise in QRC as well as in RF-QRC, and to present a few methods to mitigate its effects. Because of the lack of recurrence inside the reservoir (i.e.~ the parameterized quantum circuit), RF-QRC suppresses the propagation of correlations arising from noisy expectation values over time. RF-QRC also contains leaky integrated neurons, which introduce temporal memory and provide exponential smoothing of noisy states \citep{lukosevicius_practical_2012}. This makes RF-QRC a promising candidate for succeeding with learning tasks on noisy NISQ devices. 

This paper is structured as follows. In Sec.~\ref{sec:RC}, we provide a brief overview of classical reservoir computing with leaky integrated neurons. In Sec.~\ref{sec:RFQRC} we outline Recurrence-free Quantum Reservoir Computing (RF-QRC) as introduced in \citep{ahmed2024prediction} and compare it with leaky integrated reservoir computing without recurrence. We then extend this analysis to model uncorrelated noise in RF-QRC. Sec.~\ref{sec:denoised} compares QRC and RF-QRC with finite sampling noise and we present two denoising methods based on singular-value decomposition (SVD) and signal-smoothing techniques. These denoising methods are then applied to the three-dimensional Lorenz-63 and a nine-dimensional turbulent shear flow models (Appendix ~\ref{sec:AppL63},\ref{sec:App_MFE}). The proposal of denoising is then tested on noisy quantum hardware in Sec.~\ref{sec:hardware}. Finally, in Sec.~\ref{sec:conclusion} we conclude our findings and present prospects of future work. 

\section{\label{sec:RC}Reservoir Computing Formalism}

Reservoir computing is a type of Recurrent-Neural Network (RNN) that learns temporal correlations from the input data by mapping the low-dimensional input data to a high-dimensional reservoir. A particular type of classical reservoir computing, also known as Echo State Network \citep{jaeger__2001}, uses randomly generated input and reservoir weight matrices $\pmb{W}_{in} \, \epsilon \,\, \mathbb{R}^{{N}_{r} \, \times {N}_{u}} $ and $\pmb{W} \, \epsilon \,\, \mathbb{R}^{{N}_{r} \, \times {N}_{r}}$ to generate reservoir activation states ($\pmb{r}({t_{i+1}}) \, \epsilon \,\, \mathbb{R}^{{N}_{r}} $) at each time step from the input data $\pmb{u}_{in}(t_{i}) \, \epsilon \,\, \mathbb{R}^{{N}_{u}}$, as
\begin{align}
\pmb{\hat{r}}(t_{i+1}) &= \tanh \,(\pmb{W}_{in}  \pmb{\hat{u}}_{in}(t_{i+1})+\pmb{W} \pmb{r}(t_{i})), \label{eq:1a} \\
\pmb{r}(t_{i+1}) &= (1-\epsilon) \, \pmb{r}(t_{i}) + \epsilon \, \pmb{\hat{r}}(t_{i+1}),\label{eq:1b}
\end{align}
where $\epsilon$ is the user-defined leak-rate, which combines previous state information with the current time-step as a linear combination - also known as leaky-integral echo state network \citep{JAEGER2007335}. The $\text{tanh}$ function is applied component-wise.
\begin{figure}[!ht]
\includegraphics[width=0.9\linewidth]{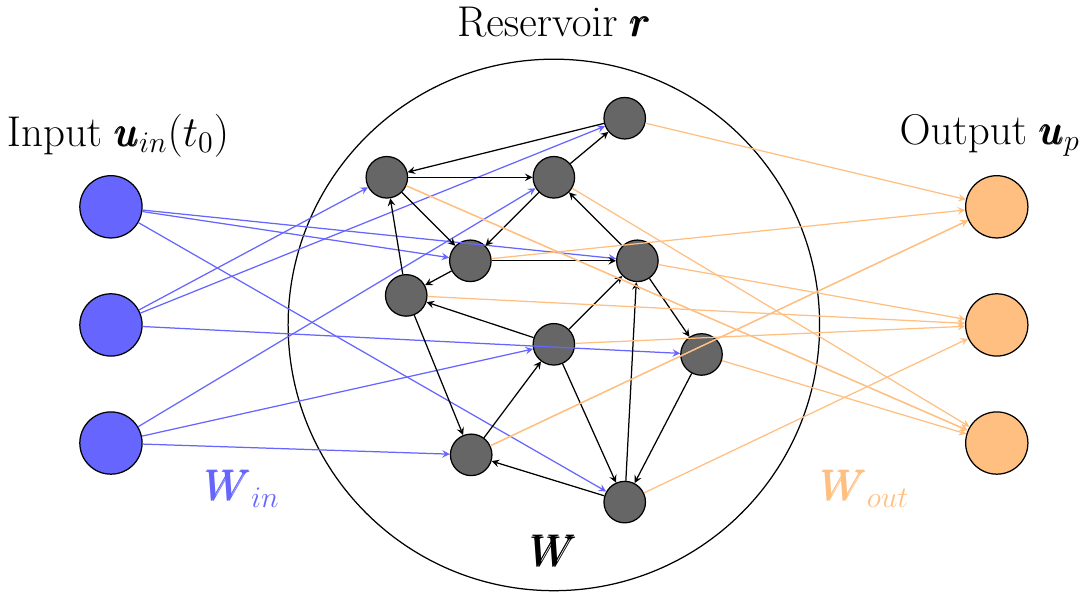}% 
\caption{\label{fig:C_Reservoir}Schematic representation of a reservoir computer \citep{jaeger__2001}. The input data $\pmb{u}_{in}$ is mapped to the reservoir matrix via $\pmb{W}_{in}$. The reservoir neuron connections governed by $\pmb{W}$ matrix allow the flow of information between neurons. The linear readout layer using the trained $\pmb{W}_{out}$ matrix is used to make output predictions $\pmb{u}_{p}$.}
\end{figure}

\begin{figure*}
    \centering
    \includegraphics[width=0.98\linewidth]{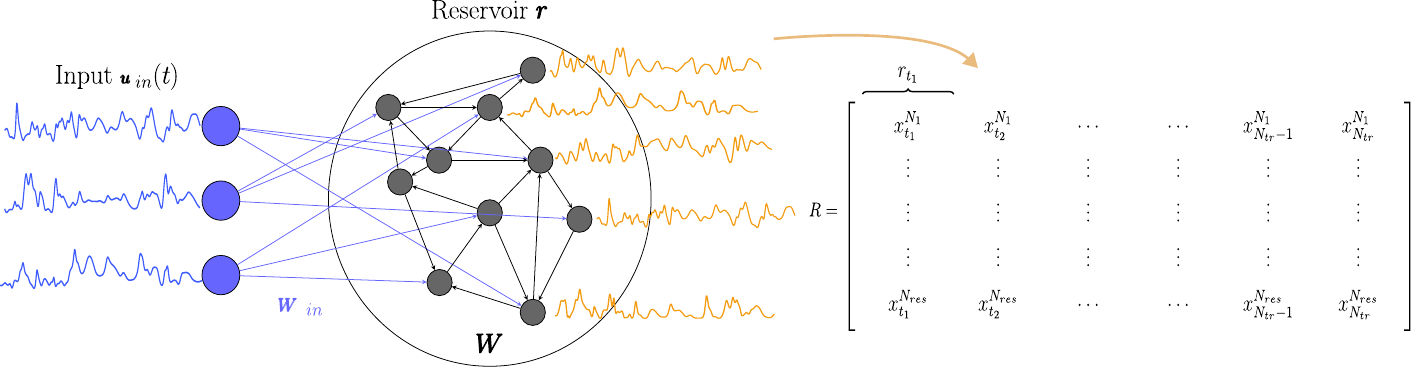}
    \caption{Training phase in reservoir computing. Input time series $\pmb{u}_{in}(t)$ is mapped to a reservoir using $\pmb{W}_{in}$ matrix for CRC and suitable encoding schemes for QRC. Inside reservoir $r$, each neuron \textit{echoes} with the input time series to generate a series of reservoir activation state signals. The reservoir activation state signals are concatenated in the reservoir state matrix $\pmb{(R)}$, which is then used for finding optimal output weight matrix $\pmb{W}_{out}$ in RC training using ridge regression.}
    \label{fig:QRC_states}
\end{figure*}

The optimal weight matrix $\pmb{W}_{out}$ is obtained by linear ridge regression 
\begin{equation}\label{eq:2}
(\pmb{R}\pmb{R}^{T}+\beta\pmb{I}) \, \pmb{W}_{out} = \pmb{R} \, \pmb{U}^{{\textnormal T} }_{d}.
\end{equation}
Here, $\pmb{R} \equiv [\pmb{r}(t_{1}), \pmb{r}(t_{2}), ... \pmb{r}(t_{N_{tr}})] \in \mathbb{R}^{{N}_{r} \, \times N_{tr}}$ is a matrix of concatenated reservoir activation signals corresponding to each neuron for $N_{tr}$ time steps of training (Fig.~\ref{fig:QRC_states}), $\beta$ is the Tikhonov regularization factor and $\pmb{U}_{d} \equiv [\pmb{u}_{in}(t_{1}), \pmb{u}_{in}(t_{2}), ... \pmb{u}_{in}(t_{N_{tr}})] \in \mathbb{R}^{{N}_{r} \, \times N_{tr}}$ is the matrix of concatenated input time series data used for training.

To predict future time series elements $\pmb{u}_{p}(t_{i+1})$, the reservoir computer can either be run in open-loop or autonomous (closed-loop) configurations  \citep{ahmed2024prediction}. The prediction, $\pmb{u}_{p}$, is a linear combination of reservoir states 
\begin{equation}\label{eq:4}
\pmb{u}_{p}(t_{i+1}) = [\pmb{r}(t_{i+1})]^{{\textnormal T} } \, \pmb{W}_{out}. 
\end{equation}

\subsection{Leaky integrated reservoir computing}\label{sec:leak}

The continuous-time dynamics of a dynamical system with leaky integration can be written as
\begin{align}\label{eq:li1}
    \pmb{x}(t) + \tau \frac{d\pmb{x}}{dt} = \pmb{F}(\pmb{x}(t),\pmb{u}(t)), 
\end{align}
where, $\pmb{x}(t)$ is the state of the dynamical system, $\tau$ is a time-constant of the system determining the rate of leakage or decay, and $\pmb{F}$ is a non-linear function describing the evolution of state $\pmb{x}(t)$ that is influenced by an input signal $\pmb{u}(t)$. For echo state networks, the function $\pmb{F}$ depends on random input and reservoir weight matrices \citep{jaeger__2001}.
\begin{align}\label{eq:li2}
    \tau \frac{d\pmb{x}}{dt} = -\pmb{x}(t) + \tanh \,(\pmb{W}_{in}  \pmb{{u}}_{in}(t)+\pmb{W} \pmb{x}(t)). 
\end{align}
% \begin{widetext}
Using Euler discretization with stepsize $\Delta t$, we obtain a discrete-time reservoir state update equation. In practical implementations, the input data $\pmb{{u}}_{in}(t)$ is also sampled at discrete time steps.
% $\pmb{{u}}_{in}(t) \equiv \pmb{{u}}_{in}(t \Delta)$ 

\begin{multline}\label{eq:li3}
    \pmb{x}(t+\Delta t) = ( 1- \frac{\Delta t}{\tau} ) \, \pmb{x}(t) + \\ \frac{\Delta t}{\tau} \tanh \,(\pmb{W}_{in}  \pmb{{u}}_{in}(t+\Delta t)+\pmb{W} \pmb{x}(t))
\end{multline}

Simplifying Eq.~\eqref{eq:li3} and keeping $\Delta t \equiv 1$ for discrete time steps, gives the same reservoir state update equations shown previously in Eqs.~\eqref{eq:1a}-\eqref{eq:1b}. Here, $\epsilon \leq 1$ is a hyperparameter governing how much information from the previous reservoir states is retained 

\begin{align}\label{eq:li4}
    \pmb{{x}}(t+1) &= ( 1- \epsilon ) \, \pmb{x}(t) + \epsilon \tanh \,(\pmb{W}_{in}  \pmb{{u}}_{in}(t+1)+\pmb{W} \pmb{x}(t))  
\end{align}

This type of model is also known as the Leaky-integrate and Fire (LIF) Neurons, which has several applications in Neuroscience \citep{teeter2018generalized,GABBIANI2010143} and signal processing \citep{lansky2008review}. The concept of leakage in reservoir computing was introduced in liquid state machines \citep{10.1162/089976602760407955} and echo state networks \citep{JAEGER2007335,lukosevicius_practical_2012}. Physically, these leaky integrated neurons have individual state dynamics that make them suitable for temporal learning tasks. Leaky integration can also be considered as a digital low-pass filter or exponential smoothing \citep{lukosevicius_practical_2012}, making the model appropriate for learning with noisy data.
\section{Recurrence-free quantum reservoir computing}\label{sec:RFQRC}

To explain the Recurrence-free quantum reservoir computing (RF-QRC) \citep{ahmed2024prediction}, we introduce standard gate-based QRC architectures. At each time step, in gate-based quantum reservoir computing the quantum state vector is propagated by a $\pmb{\theta}$-parametrised quantum circuit with unitary $\mathcal{U}(\pmb{\theta}$)
\begin{equation}\label{eq:5}
|\psi(t_{i+1}) \rangle  = \mathcal{U} (\pmb{\theta}) |\psi(t_{i}) \rangle .
\end{equation}
Previous proposals of gate-based QRC \citep{pfeffer_hybrid_2022} involve a reservoir map that depends on input data $\pmb{u}_{in}$, previous reservoir states $\pmb{r}$ and a random unitary parameterized by $\pmb{\alpha}$. Specifically,
\begin{equation}\label{eq:6}
   {|\psi(t_{i+1}) \rangle } =  \mathcal{U}(\tilde{\pmb{\alpha}}) \mathcal{U}(\pmb{u}_{in}(t_{i+1})) \mathcal{U}( \pmb{r}(t_{i}))  | 0 \rangle^{\otimes n}.
\end{equation}
Equation~\eqref{eq:6} is analogous to the classical reservoir update in Eq.~\eqref{eq:1a}. 

By contrast, the Recurrence-free Quantum reservoir state updating equation only evolves as a function of the input time series and a random $\boldsymbol{\alpha}$-parametrised unitary.
\begin{equation}\label{eq:7}
   {|\psi(t_{i+1}) \rangle } =  \mathcal{U}(\pmb{\alpha}) \mathcal{U}(\pmb{u}_{in}(t_{i+1}))| 0 \rangle^{\otimes n}.
\end{equation}

After each time stepping, a measurement in the computational basis $\{|k\rangle\}_{k=0}^{k=2^n}$ is performed and the result is used to form a new reservoir state vector $\boldsymbol{r}(t_{i+1})$ 
\begin{equation} \label{eq:8} 
    {r}^{(k)}(t_{i+1}) = (1-\epsilon) \, r^{(k)}(t_{i}) + \epsilon \, |\langle \psi(t_{i+1}) | k\rangle| ^2
\end{equation}
Equation~\eqref{eq:8} resembles the leaky-integrated quantum reservoir state update with recurrence Eq.~\eqref{eq:li4}, but is different in that the state update in the parameterized quantum circuit only depends on input data $\pmb{u}_{in}$ and not on the recurrent $\pmb{r}(t_{i})$ at each time step.

The RF-QRC may resemble a Quantum Extreme Learning Machine (QELM) because it does not have an active recurrence inside the reservoir (parameterized quantum circuit) \citep{mujal_opportunities_2021,xiong2023fundamental}. However, RF-QRC is not a QELM because it also contains information about previous reservoir states using leaky-integrated neurons with hyperparameter $\epsilon$. It has been observed that $\epsilon$ can be tuned to alter the short-term memory capacity \citep{lukosevicius_practical_2012}. This also makes RF-QRC suitable for temporal learning tasks as discussed in ~\citep{ahmed2024prediction} for various chaotic systems. Additionally, the performance of RF-QRC depends on the chosen feature map which is analogous to the necessity of tuning classical reservoirs in classical reservoir computing. 
Owing to the lack of active recurrence, RF-QRC is a specific kind of quantum reservoir computer that is scalable and potentially offers a natural path to mitigate the issue of propagating noise.

\subsection{Quantum reservoir computing with finite samples}

\begin{figure}
    \centering
    \includegraphics[width=1.0\linewidth]{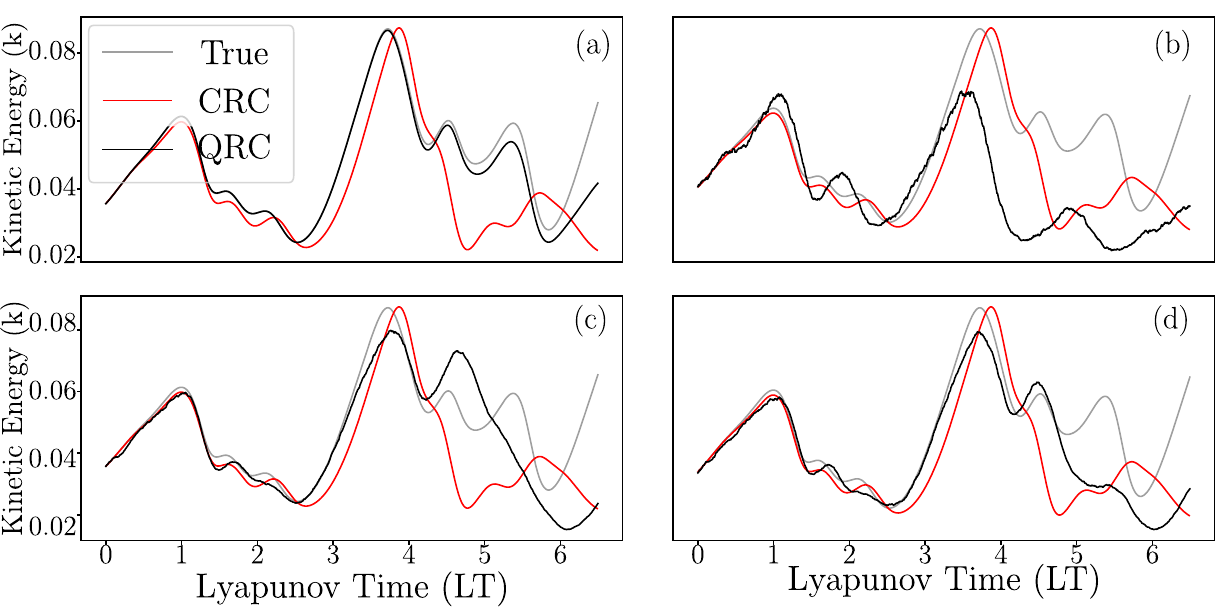}
    \caption{Prediction of the MFE time-series with (a) noise-free probability distribution (b) 0.5$\times 10^5$ shots (c) 2$\times 10^5$ shots (d) 4$\times 10^5$ shots.}
    \label{fig:QRC_MFESHOTS}
\end{figure}

In this section, we study the impact of finite sampling noise on the prediction capabilities of QRC and RF-QRC. In Fig.~\ref{fig:QRC_MFESHOTS}, the MFE \citep{moehlis_low-dimensional_2004}  time-series prediction is shown for an ideal probability distribution (assuming an infinite number of measurements, sometimes referred to as `shots'), and for various different learning outcomes based on a variable number of shots $S$. These results indicate that a certain minimum number of finite samples is required to improve the forecasting abilities of QRC beyond classical reservoir computers. 

Let us now consider more specifically the effects of finite-sampling noise in RF-QRC, which because of the  absence of active recurrence can generally be assumed as noise uncorrelated in time. %For $n$ qubits, the state vector dimension and thus the total reservoir size is $N_{res} = 2^{n}$. 
When neglecting hardware noise, the actual reservoir state $\boldsymbol{{r}}_N(t)$ is related to the outcomes of $S$ measurements $\overline{\boldsymbol{r}}_N(t)$ by\footnote{We will drop the subscript from $t_{i+1}$ for brevity.} 
\begin{align}\label{eq:9} 
\pmb{{r}}_{N}(t) = \lim_{S\to\infty} \overline{\pmb{{r}}}_{N}(t).
\end{align}
The effect of finite sampling noise on reservoir activation states $(\pmb{r}_{N})$ can be modeled as a time-dependent stochastic variable $\boldsymbol{\zeta}(t)$ with an explicit constant prefactor of $1/\sqrt{S}$ accounting for the central limit theorem. 
\begin{align}\label{eq:10} 
\overline{\pmb{{r}}}_{N}(t) =  {\pmb{{r}}_{N}(t) }+ \frac{1}{\sqrt{S}} \boldsymbol{\zeta}(t).
\end{align}
Like in classical reservoir computing, in QRC, computational basis states correspond to neurons $N_{1}, N_{2},..., N_{res}$ where $N_{res}=2^n$ is the dimension of the reservoir. The time varying shot noise signals may therefore be regarded as time dependent noise functions on those neurons as illustrated in Fig.~\ref{fig:QRC_states}.
The stochastic nature of individual reservoir vectors translates to a reservoir state matrix $R$ with stochastic variability $\boldsymbol{Z}$
\begin{align}\label{eq:11} 
\overline{\pmb{{R}}} =  {\pmb{{R}} }+ \frac{1}{\sqrt{S}} \, \pmb{Z}.
\end{align} %(\pmb{{R}}_{N})

Quantum sampling noise of a single qubit follows a binomial distribution \citep{schuld2021machine}. For multiple qubits forming a quantum state vector with finite samples the sampling noise becomes a multinomial distribution. The lack of recurrence in RF-QRC allows us to model this noise as uncorrelated in time. This form of noise has also been considered in \citep{hu2023tackling} in the context of the analysis of Resolvable Expressive Capacity (REC).

Consequently, in Eq.~\eqref{eq:11}, we can model ${\pmb{Z}}$ as a centered multinomial stochastic process. Without loss of generality, ${\pmb{Z}}$ can always be shifted to have zero mean ($\mathbb{E} \, [{\pmb{Z}}] =  0$). Thus, the stochastic matrix can be modeled by only considering second-order moments that form the sampled covariance matrix 
\begin{align}\label{eq:12} 
{\pmb{\Sigma}_{ij}(t)} = Cov[\zeta_{i}(t),\zeta_{j}(t)]. 
\end{align}
By taking the expectation value over an infinite number of measurements one gets the covariance matrix $ \boldsymbol{V} = \mathbb{E} \, [{\pmb{\Sigma}_{ij}}]$. It can be written in terms of the sampled reservoir matrix $\boldsymbol{R}$
\begin{align}\label{eq:13} 
 \boldsymbol{V} = {\rm diag}\left(\frac{1}{N_{tr}}\sum_{N_{tr}}(\overline{\pmb{{R}}})\right) - \overline{\pmb{{R}}}\,\overline{\pmb{{R}}}^{T}.  
\end{align} 
The second term on the right-hand side of Eq.~\eqref{eq:13}, also known as Gram Matrix, arises naturally in the ridge-regression loss in reservoir computing Eq.~\eqref{eq:2}. This type of loss function is quadratic and cumulants up to second-order (mean and covariances) are generally assumed to be sufficient to study the effect of noise in the training of these models \citep{hu2023tackling,khan2023practical}. 

\section{Noisy reservoir activation states}\label{sec:denoised}
The working principle of reservoir computing is to combine the generated reservoir activation signals for functional approximation of the dynamical systems (Fig.~\ref{fig:QRC_states}). This is done by minimizing the loss function over the input training data set. Reservoir signals have a large overlap with a relatively lower-dimensional manifold \citep{carroll2020dimension} of active states, which is the active space. More specifically, the singular value decomposition of common reservoir matrices reveals only a limited number of relevant singular values and any denoising procedure should preserve the corresponding eigenspaces.

For a number of chaotic systems, the time series signals have corresponding reservoirs with low dimensional active spaces \citep{carroll2020dimension}. Often, in the classical reservoir computing framework, reservoir matrices with high-dimensional active spaces are associated with lower testing errors \citep{carroll2019network}. However, in QRC we find that, although the addition of finite sampling noise in QRC reservoirs increases the dimensionality of the active space, it does not necessarily improves the learning performance. REC analysis \citep{hu2023tackling} shows that in the presence of noise taking more activation states than a given threshold results in poor functional approximation. 
Therefore, in order to perform an analysis of the impact of noise on the learning performance, we consider the signal-to-noise (SNR) ratios of individual reservoir activation signals. Furthermore, we compute the accuracy of function fitting of these noisy signals, which is determined by computing the training loss.

\begin{figure}
    \centering
    \includegraphics[width=0.75\linewidth]{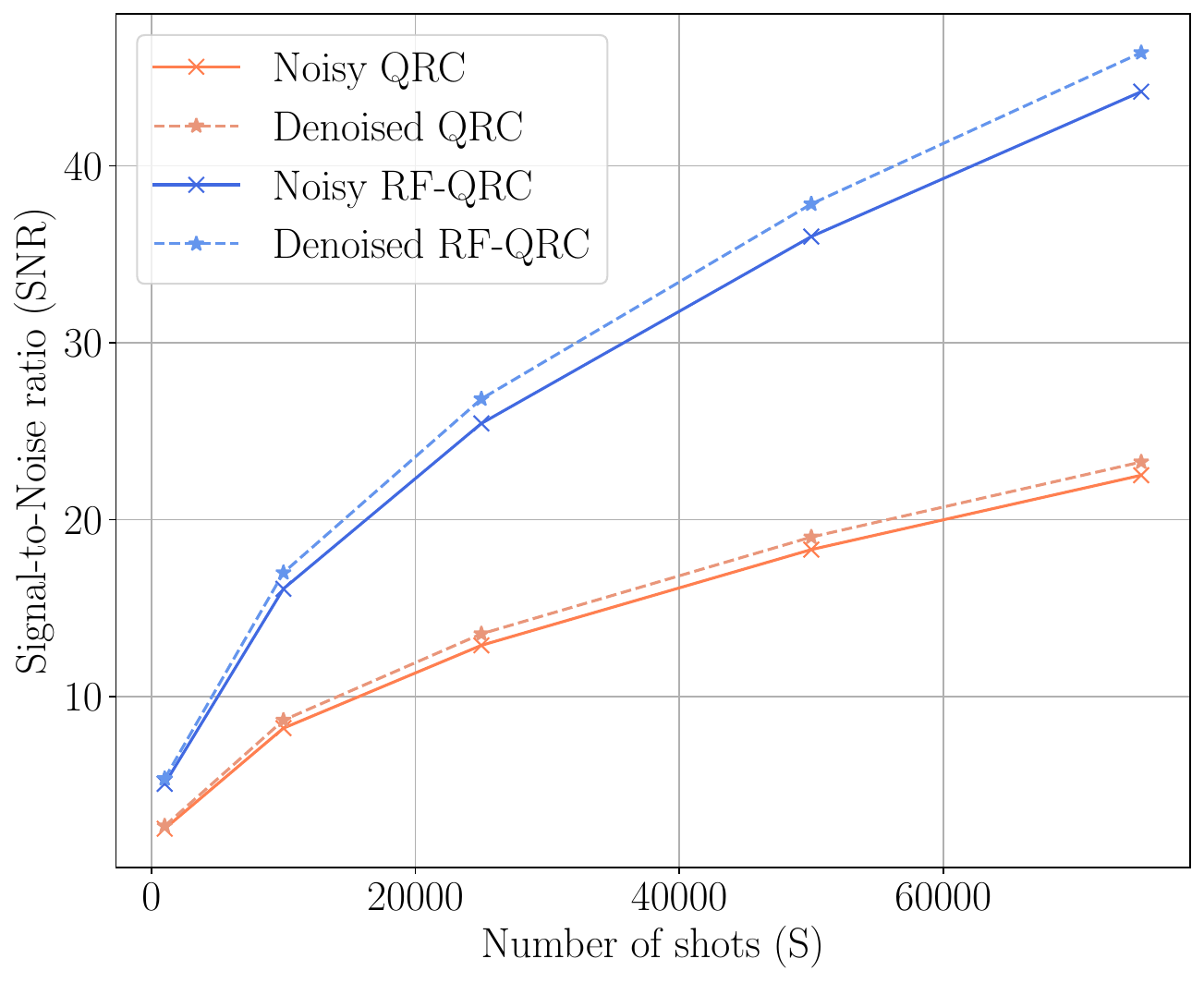}
    \caption{Averaged signal-to-noise ratios (SNR) of the reservoir states of QRC and RF-QRC with and without denoising filters. Underlying model is an MFE times series.}
    \label{fig:MFE_SNR_R}
\end{figure}

\begin{figure}
    \centering
    \includegraphics[width=0.99\linewidth]{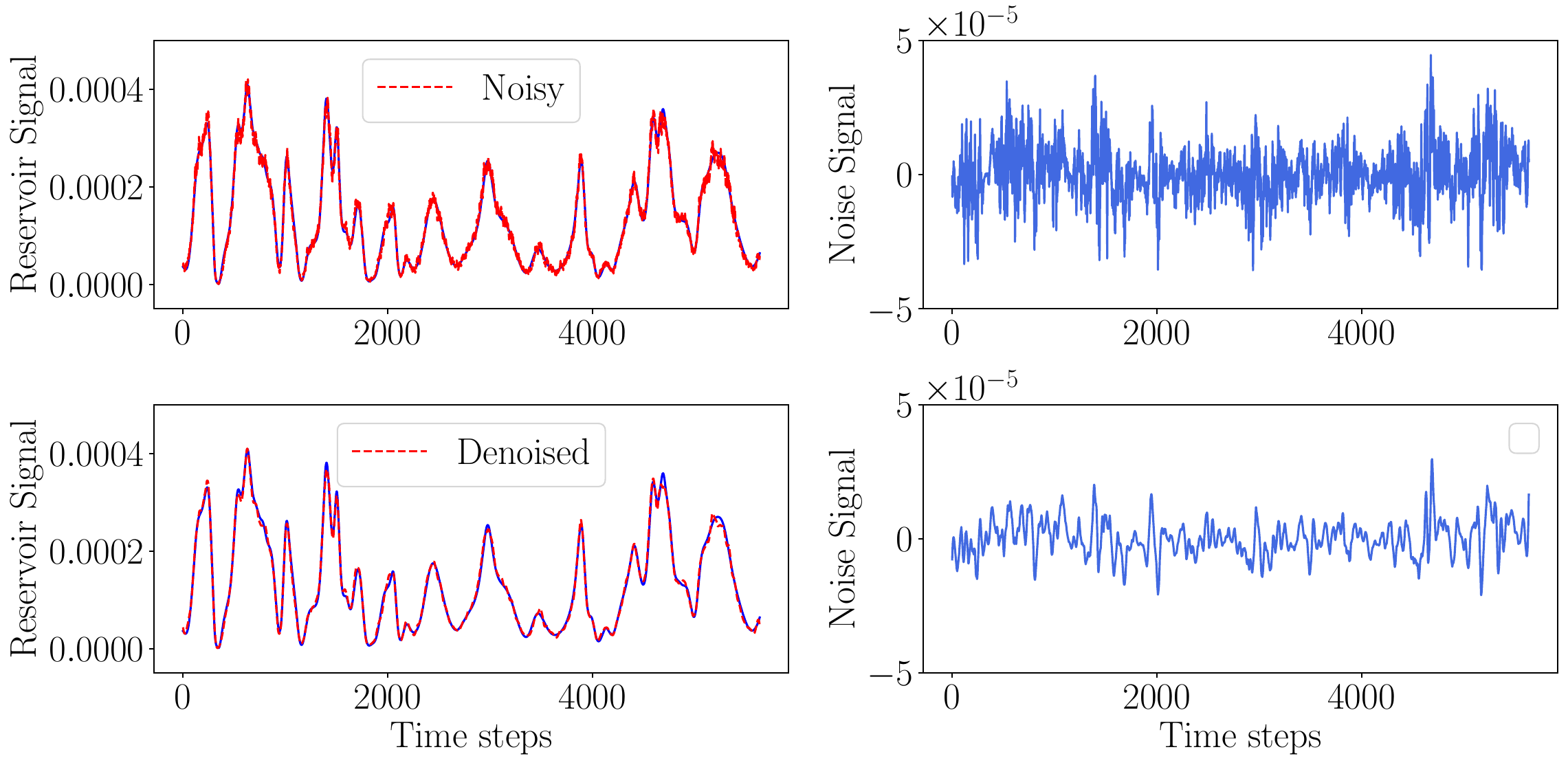}
    \caption{Example of reservoir activation state time series with 50k shots at each timestep for RF-QRC of the MFE system. Top, visualization of the noisy reservoir activation signal and its noise component; bottom, visualization of the denoised reservoir activation signal and its noise component.}
    \label{fig:MFE_visual}
\end{figure}

In Fig.~{\ref{fig:MFE_SNR_R}}, we compare the SNR ratio for QRC and RF-QRC for a 9-dimensional chaotic shear flow model of the MFE system \citep{moehlis_low-dimensional_2004}. Our results indicate that the presence of correlated noise in QRC results in more noisy estimates of the reservoir activation signals than RF-QRC because of the propagation of correlations in QRC. An example of the noisy and denoised reservoir activation state signals for RF-QRC, measured on a 10 qubits system, can be seen in Fig.~{\ref{fig:MFE_visual}}. Our results indicate that for a constant number of samples denoising reservoir activation signals results in a better fit of the training signal with lower noise variance. We now discuss the details of the two approaches utilized for suppressing noise in RF-QRC.

\subsection{Noise suppression using SVD}
In signal processing, principal component analysis (PCA) or singular value decomposition (SVD) is a method to improve the SNR of noisy signals 
\citep{shlens2014tutorial,schanze2017removing,jha2010denoising}. The unbiased estimation of the expectation values in quantum computation is fundamentally limited by the Cramer-Rao bound \citep{yu2022quantum}. For an ensemble of quantum systems, governed by an input time series, the resulting expectation values form a reservoir signal with an added finite-sampling noise. We found that the SNR of these noisy reservoir signals can be improved by using classical signal processing tools such as SVD. For a noisy reservoir state matrix $\overline{\pmb{R}} \, \in \, \mathbb{R}^{N_{r} \times N_{tr}} $, the singular value decomposition is
\begin{align}\label{eq:14} 
 \overline{\pmb{R}} = \pmb{U} \, \pmb{S} \, \pmb{V}^{T} 
\end{align}
where $\textbf{U}$ is an orthogonal $N_{r} \times N_{tr}$ matrix,  $\textbf{S}$ is a diagonal ${N_{tr} \times N_{tr}}$ matrix with non-negative singular values, and  $\textbf{V}$ is an orthogonal ${N_{tr} \times N_{tr}}$ matrix. In order to maximize the SNR, we derive a low-rank approximation of our noisy reservoir matrix by truncating singular values below a given threshold that are assumed to be associated with the noise component.

\begin{figure}
    \centering
    \includegraphics[width=0.99\linewidth]{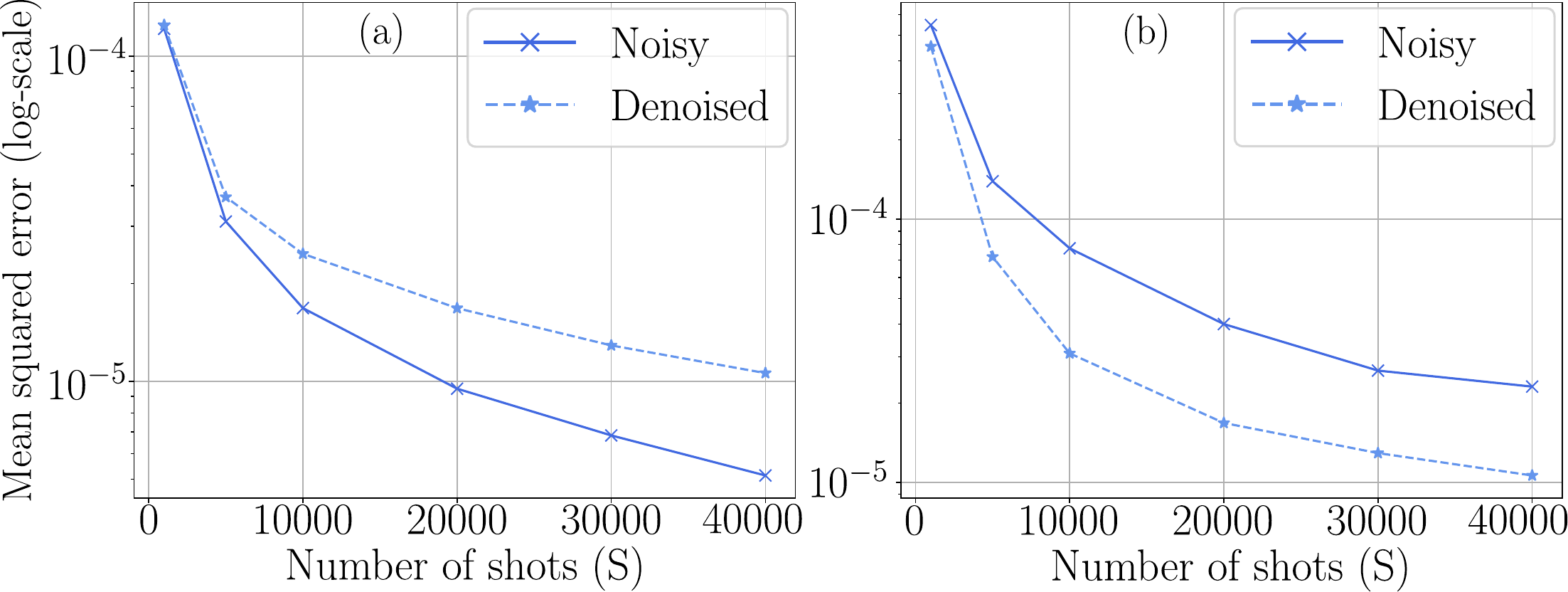}
    \caption{Time series prediction of the Lorenz 63 model using RF-QRC. Averaged mean-squared errors of the trajectory are shown as a function of the number of measurements $S$ for noisy and SVD denoised protocols.  a) Comparison of noisy and denoised figures with an underlying original reservoir size of $N_{res} = 128$. b) Comparison of noisy and denoised figures with an underlying reduced reservoir size of $N_{res} = 60$.}
    \label{fig:L63_SVD_2}
\end{figure}
\begin{figure}
 \centering
 \includegraphics[width=1.0\linewidth]{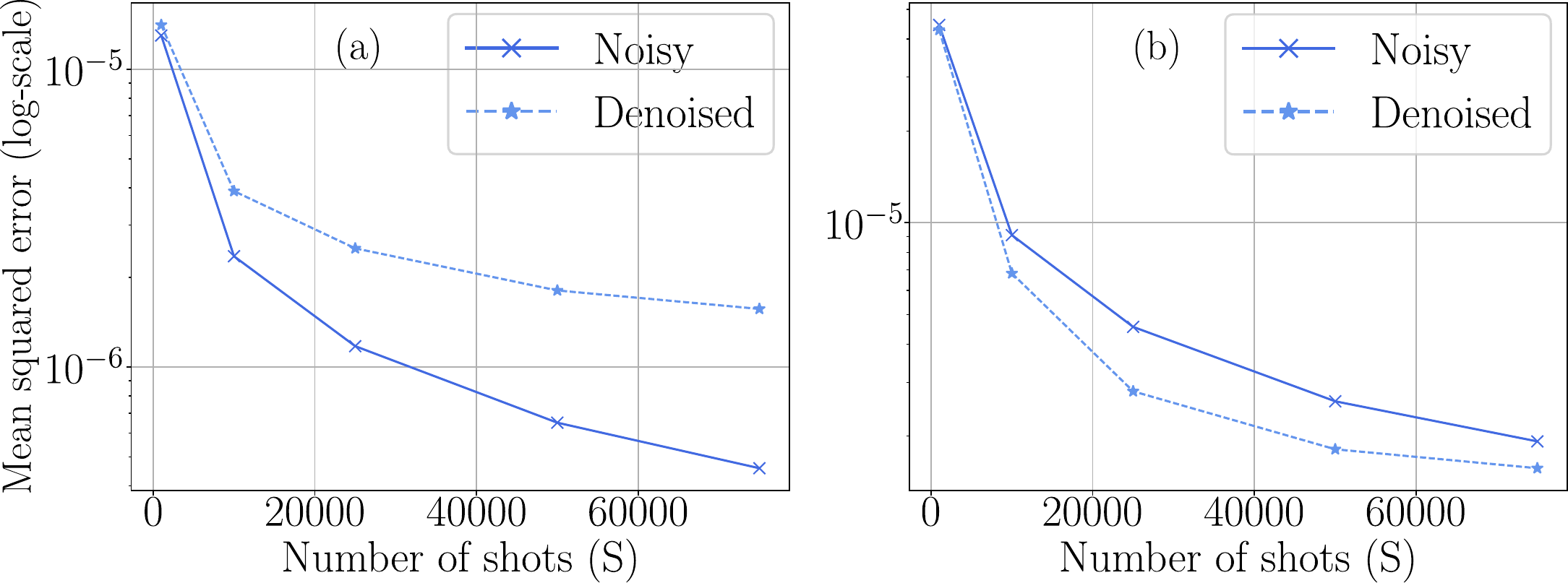} 
  \caption{Time series prediction of the MFE model using RF-QRC. Averaged mean-squared errors of the trajectory are shown as a function of the number of measurements $S$ for noisy and SVD denoised protocols.  a) Comparison of noisy and denoised figures with an underlying original reservoir size of $N_{res} = 1024$. b) Comparison of noisy and denoised figures with an underlying reduced reservoir size of $N_{res} = 500$.}
 \label{fig:MFE_SVD_2}
\end{figure}

In Fig.~\ref{fig:L63_SVD_2}, we show the results of the training error for a 7-qubits (${N}_{res}=128$) reservoir size trained on a Lorenz-63 time series with finite quantum reservoir sampling. We compare the results of the noisy estimates with the denoised estimates obtained by performing SVD of the noisy reservoir matrix. These results show that obtaining a low-rank approximation of the reservoir matrix improves the training error when compared to a reduced noisy reservoir matrix of the same size (Fig.~\ref{fig:L63_SVD_2}, b). However, for the complete reservoir representation, the mean-squared error is smaller for the original (noisy) reservoir matrix (Fig.~\ref{fig:L63_SVD_2}, a). We have found similar results for the MFE model in Fig.~\ref{fig:MFE_SVD_2}.

\subsection{Noise suppression using signal filtering}\label{sec:poly}

Using SVD for denoising requires the knowledge of the complete reservoir matrix $\pmb{R}$, the dimension of which scales exponentially with the number of qubits,  which limits the feasibility of SVD analysis for RF-QRC. In this section, we propose a second method of suppressing noise from the reservoir activation states by applying a denoising low-pass filter to each reservoir activation state. 

\begin{figure*}
    \centering
    \includegraphics[width=0.98\linewidth]{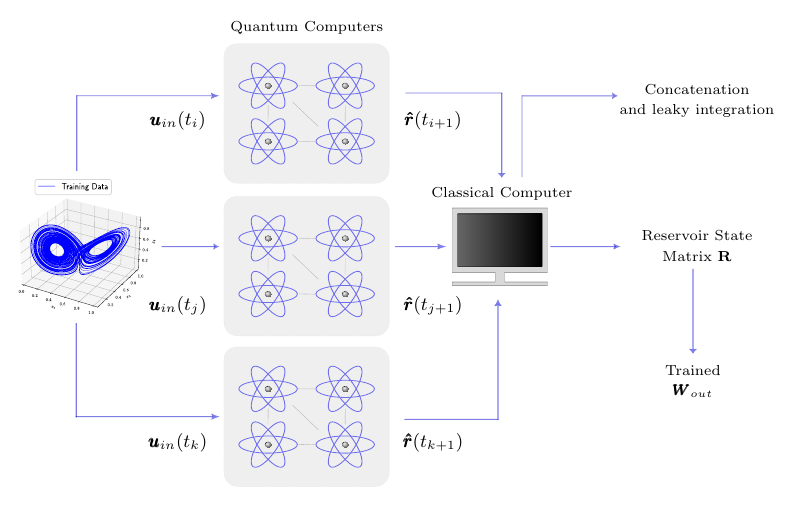}
    \caption{Recurrence-free quantum reservoir computing parallel training. Input data $\pmb{u}_{in}(t)$ is divided into smaller length signals and passed through multiple parallel quantum computers. The obtained reservoir activation signals $\pmb{\hat{r}}_{in}(t)$ associated with each input time series signal are concatenated classically where leaky integration and denoising are applied. The combined reservoir state matrix $\pmb{R}$ is then used for training using ridge-regression.}
    \label{fig:RFQRC}
\end{figure*}

We emphasize that in the case of RF-QRC and the absence of recurrence, reservoir activation states are only driven by the input time series, which are known \textit{a priori}. Thus, we can  employ multiple quantum systems in parallel to generate the reservoir activation states at each time step (Fig.~\ref{fig:RFQRC}). Later on, we can concatenate the expectation of the estimates for each eigenbasis to form a reservoir activation signal. This signal is the noisy estimate of our reservoir signal in which we remove the high frequencies by applying a \textit{moving average polynomial regression} method \citep{schafer2011savitzky}. In principle, one could also apply a physical filter to the noisy signal estimates  \citep{khan2023practical}; in this work, we employ digital filtering by post-processing quantum measurements on a classical computer. 

\begin{figure}
    \centering
    \includegraphics[width=0.98\linewidth]{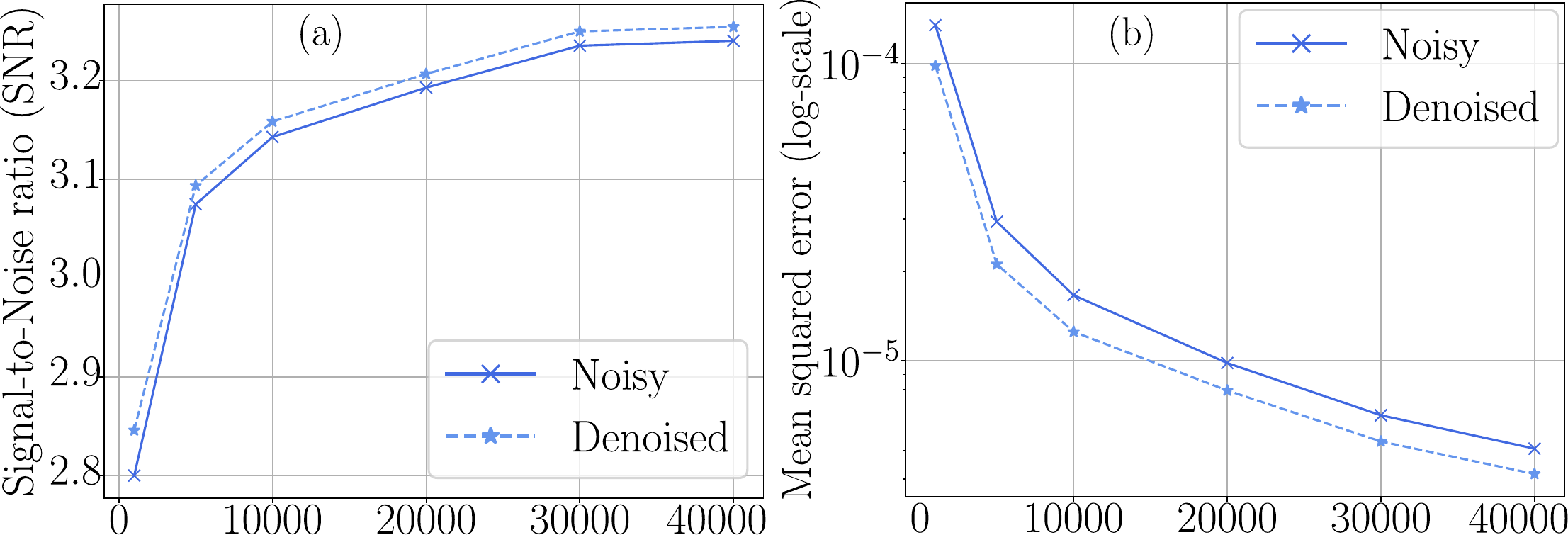}
    \caption{Comparison of (a) signal-to-noise ratios and (b) mean-squared errors of noisy and denoised RF-QRC trained on a time series of the Lorenz 63 system. Reservoir size ${N_{res} = 128}$ represented by 7 qubits. }
    \label{fig:L63_DN_2}
\end{figure}

 \begin{figure}
     \centering
     \includegraphics[width=0.9\linewidth]{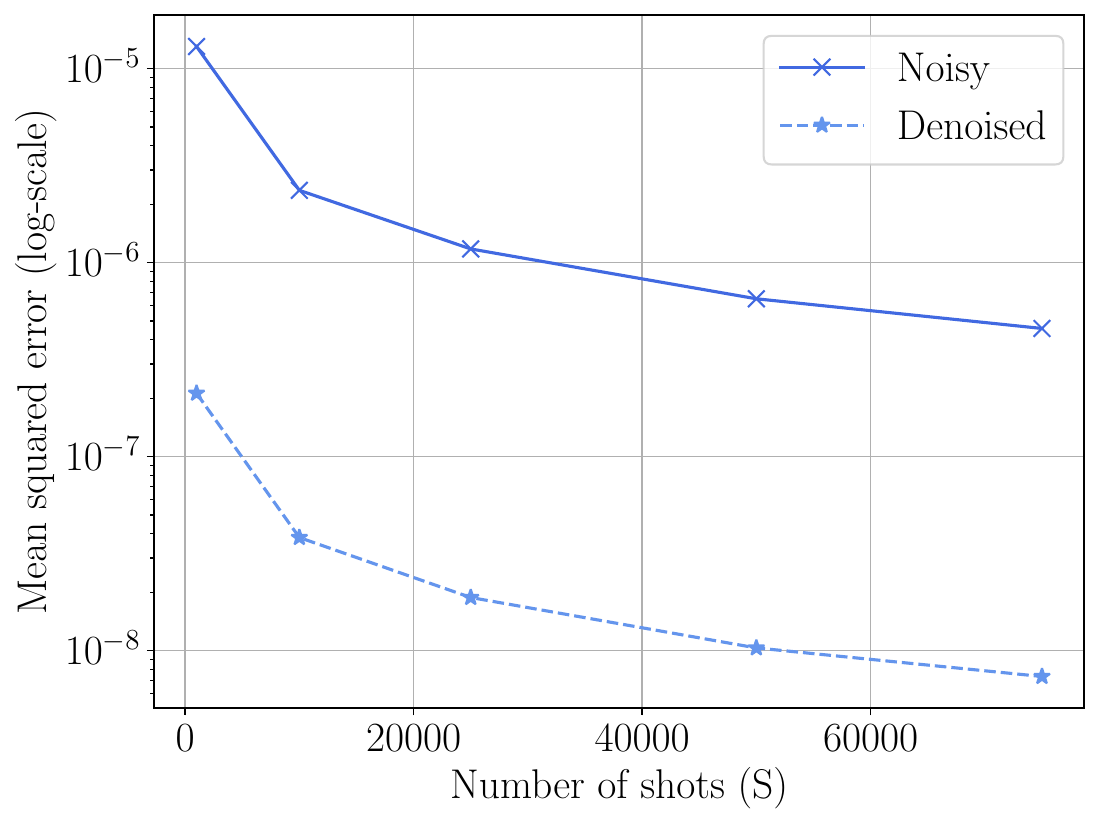}
     \caption{Turbulent chaotic shear flow time series analysis with RF-QRC, using 
     10 qubit systems. Mean-squared error for noisy and denoised activation states with signal filtering according to Sec.~\ref{sec:poly}.}
     \label{fig:MFE_DN_3}
 \end{figure}

In Fig.~\ref{fig:L63_DN_2}, for a time series of the Lorenz 63 system, we present the SNR and mean-squared training error for noisy and denoised reservoir activation states using polynomial regression. By contrast to denoising based on SVD, the results in Fig.~\ref{fig:L63_DN_2} show that using this method for suppressing noise always results in a lower mean-squared error for various numbers of shots and for increasing reservoir states. The results in Fig.~\ref{fig:L63_DN_2} also show that denoising leads to a lower mean-squared error for various numbers of shots. The lower mean-squared error therefore gives an improved SNR ratio for the reservoir activation signals. This analysis can be  extended further by employing more advanced classical filtering techniques. In Fig.~\ref{fig:MFE_DN_3}, similar results are shown for the MFE model, indicating that denoising reduces the training error significantly.

\section{Hardware Implementation}\label{sec:hardware}

\begin{figure}
    \centering
    \includegraphics[width=0.95\linewidth]{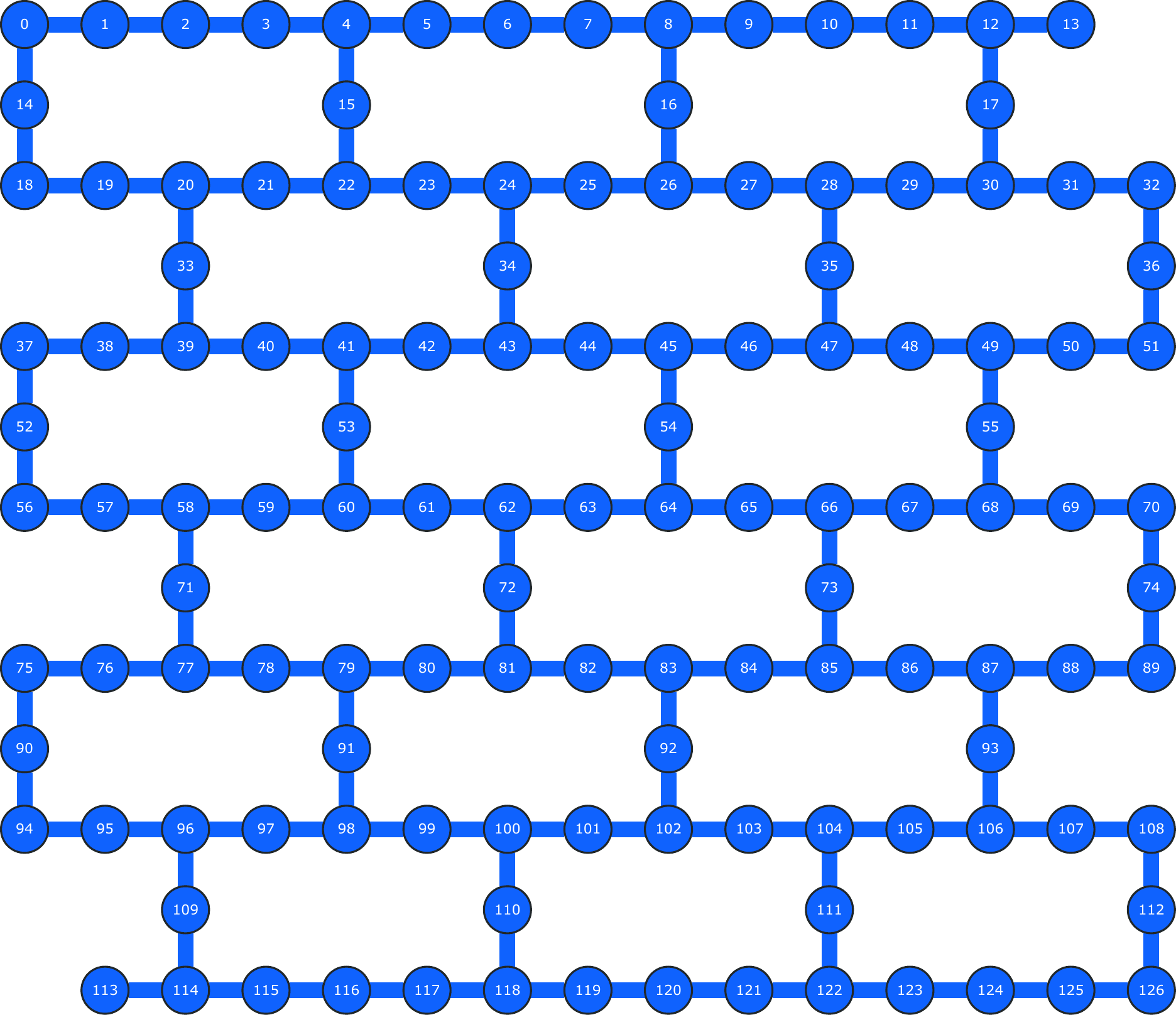}
    \caption{Connectivity and qubit map of the $\text{ibmq\_kyoto}$ 127 qubit device. 10 of the qubits with the lowest readout assignment error are used as a reservoir for  RF-QRC of the MFE model.}
    \label{fig:kyoto_map}
\end{figure}

To demonstrate the feasibility of training RF-QRC on quantum hardware and test our proposal of denoising, we train the turbulent chaotic shear flow model on the IBM Quantum 127-Qubit $\text{ibmq\_kyoto}$ device \citep{ibm}. The qubit connectivity map of the hardware backend is shown in Fig.~\ref{fig:kyoto_map}. From the 127-qubit system, we train our MFE model on a 10-qubit space that corresponds to a reservoir size of $N_{res} = 1024$. 

Before proceeding with the training of RF-QRC, we simplified the ansatz from fully connected layers to linear entangling layers \citep{ahmed2024prediction}. This simplification is necessary because the device used has limited connectivity. Our simplified RF-QRC contains a layer of Hadamard gates applied on each qubit followed by $R_y$ rotation gates, parameterized by the input time series signal (9-parameters for MFE). The circuit is then followed by a linear entangling layer of CNOT gates entangling all 10 qubits. This first feature map is applied twice to enrich reservoir dynamics. The second feature map differs from the first in that the rotation gates $R_y$ are now parameterized by random rotation angles $\pmb{V}(\alpha)$ uniformly sampled from the interval $[0,4\pi]$, to introduce randomization in the reservoir. These random rotation gates are sampled once and are kept constant throughout the training.  

The quantum circuit is then transpiled by the $Qiskit$ runtime transpiler to achieve an optimal map of the quantum circuit to the physical qubits of the $\text{ibmq\_kyoto}$ backend. Finally, dynamic decoupling \citep{Qiskit} to the idle qubits is applied to mitigate decoherence. No additional error mitigation strategies were applied.

\subsection{Parallel training and denoising on multiple QPUs}

The MFE input time series signal parameters are mapped on the pre-processed parameterized quantum circuit. The length of the time series signal was chosen as 1200 time steps. The training procedure for RF-QRC is illustrated in Fig.~\ref{fig:RFQRC}. Each quantum circuit is executed in parallel, independent of the other quantum circuits because of the lack of recurrence inside the circuit and the measurement in the computational basis is performed. For each data point, $10^4$ shots are sampled, and the results are concatenated classically, followed by leaky integration and denoising techniques as outlined in Sec.~\ref{sec:denoised}. 

\begin{figure}
    \centering
    \includegraphics[width=1.00\linewidth]{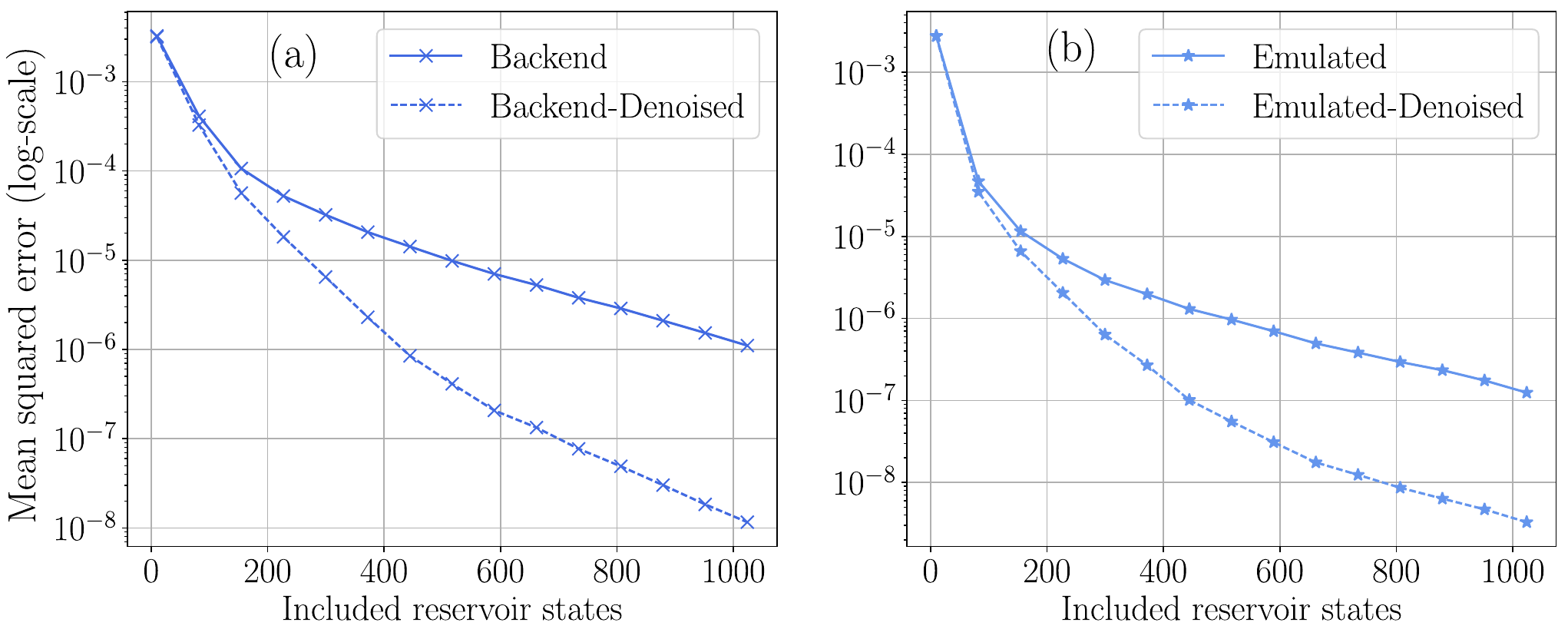}
    \caption{Turbulent chaotic shear flow (MFE) training mean-squared error comparison with noisy and denoised states for both emulated and quantum processor-based RF-QRC. }
    \label{fig:MSE_nvsdn}
\end{figure}

\begin{figure}
    \centering
    \includegraphics[width=1.00\linewidth]{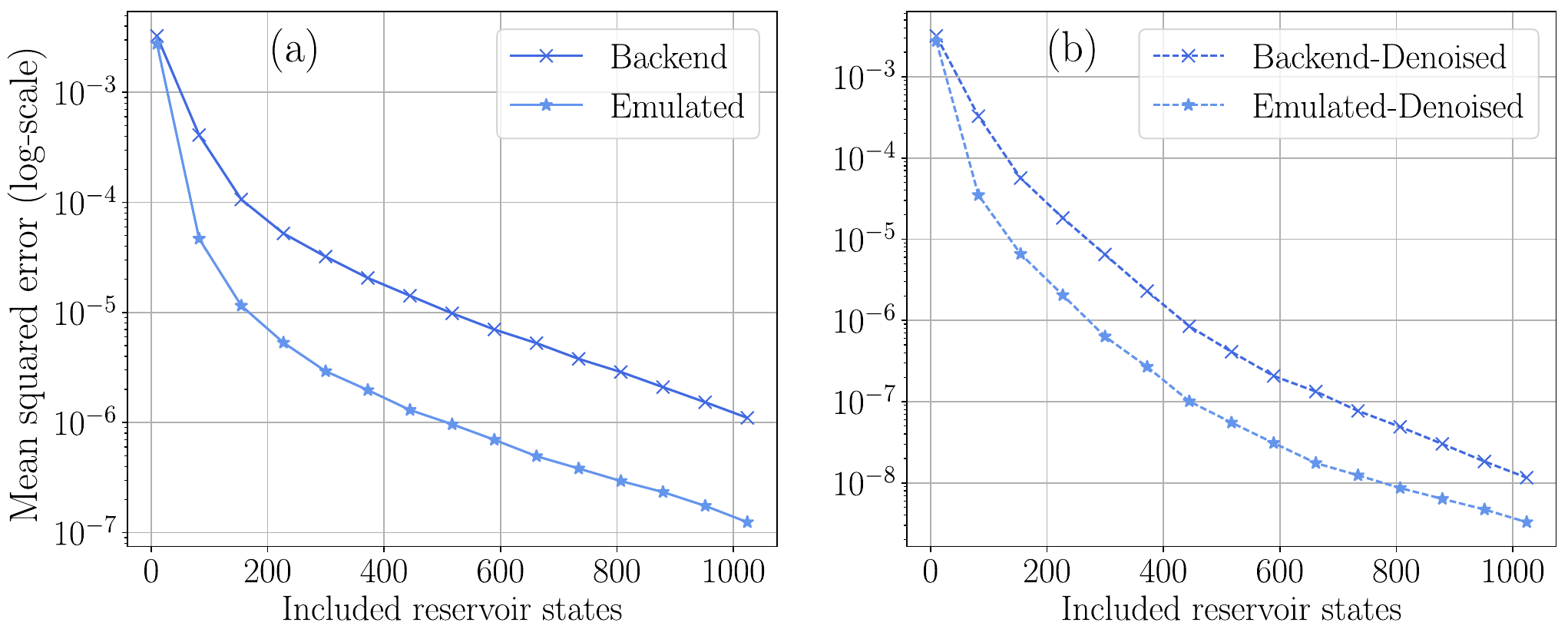}
    \caption{Turbulent chaotic shear flow training mean-squared error comparison with emulated and quantum processor results.}
    \label{fig:MSE_emulvsback}
\end{figure}

Results in Fig.~\ref{fig:MSE_nvsdn} indicate that denoising helps to improve the training accuracy for both the emulation of the state vector propagation (only subject shot noise) and the training 
on the $\text{ibmq\_kyoto}$ quantum processor (subject to shot noise \textit{and} hardware noise) for increasing reservoir states. Therefore, even in the presence of correlated hardware noises, denoising improves the performance of the model. Other results in Fig.~\ref{fig:MSE_emulvsback} demonstrate that hardware noise also has a significant impact on the performance, in addition to the impact of shot noise. In particular, the model trained on the quantum processor requires more reservoir states (i.e. a larger active space of the reservoir) for the same training accuracy when compared to a model with only shot noise. Similar effects are observed in both denoised models. However, denoising can be instrumental in mitigating those errors to reduce this gap between hardware and emulated reservoirs. This suggests that denoising suppresses shot noise, and potentially other different types of noises as well. The averaged training loss for each configuration with all included reservoir states ($N_{res} = 1024$) is listed in Tab.~\ref{tab: MFE_loss}.
\begin{table}[!htbp]
\caption{{Training loss comparison on emulator and backend trained networks, with and without denoising}}

\begin{tabular}{llll}
& Noisy & Denoised & \\ \hline
& & &\\
Emulator & 1.24 $\times 10^{-7}$  & 3.28 $\times 10^{-9}$ \\ 
Backend & 1.10 $\times 10^{-6}$ & 1.16 $\times 10^{-8}$ &  \\
& & &\\
\end{tabular}
\label{tab: MFE_loss}
\end{table}

\begin{figure}
    \centering
    \includegraphics[width=1.00\linewidth]{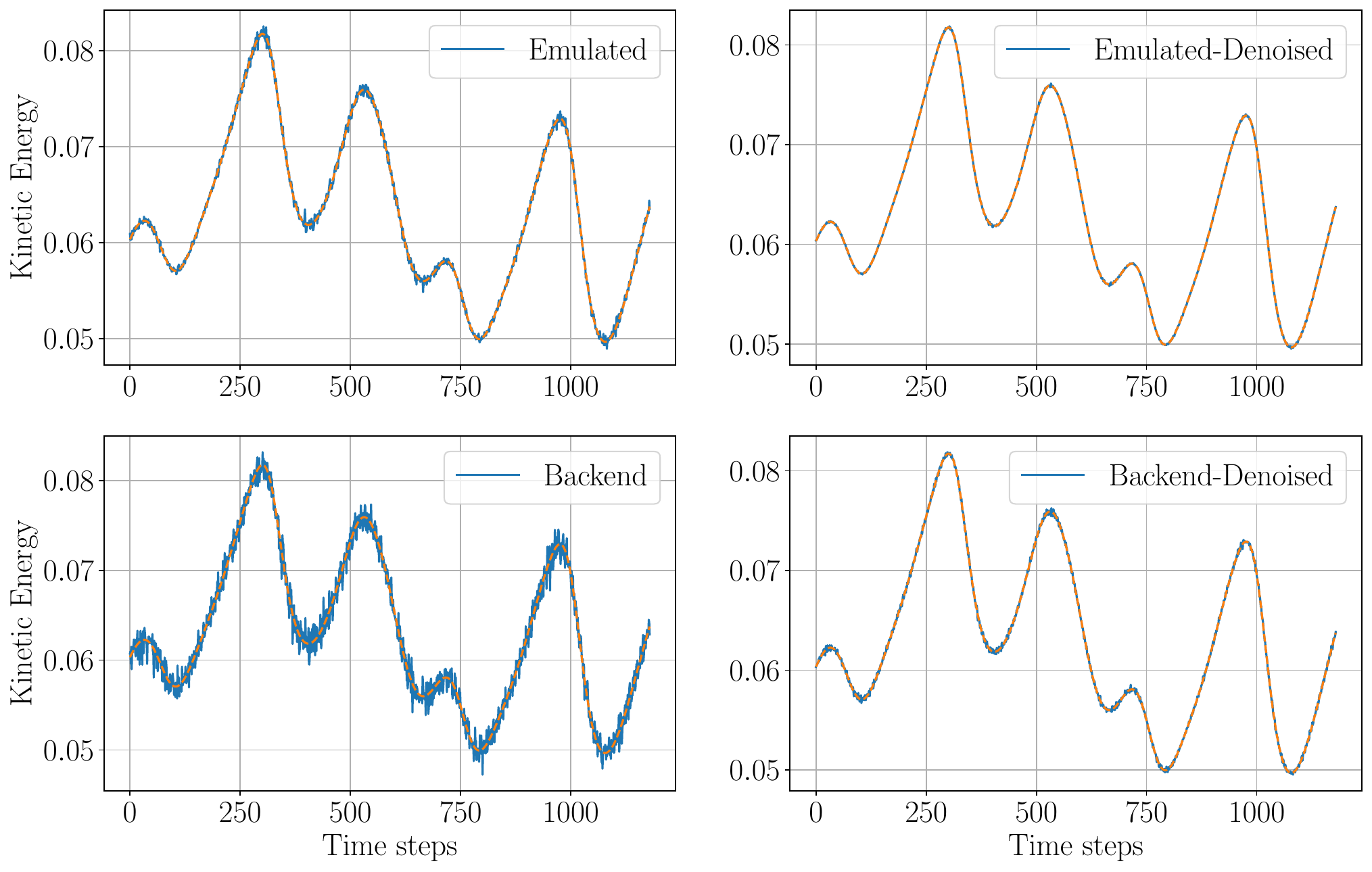}
    \caption{Turbulent chaotic shear flow reconstructed kinetic-energy comparison with emulated and backend results. The top panel presents the results of emulated noisy and denoised reservoir activation states. The bottom panel presents the results of hardware-trained noisy and denoised reservoir activation states.}
    \label{fig:MSE_KE}
\end{figure}

Finally, we present the results of the reconstructed time series of kinetic energy from a trained 9-mode MFE in Fig.~\ref{fig:MSE_KE}. We do so for both emulated models with shot noise and the trained RF-QRC on the backend including hardware noises. The reconstructed time series demonstrates that without denoising, the resulting fit for the trained signal suffers more from finite sampling and hardware noise. However, the training of RF-QRC coupled with denoising helps in reducing the shot noise as well as various hardware noises to have a better fit for optimal training. 

\section{Conclusion}\label{sec:conclusion}

Quantum reservoir computing is a promising tool for time-series forecasting of chaotic signals when emulated classically with the assumption of ideal (noise-free) expectation values. To realize any quantum advantage and for real-world applications in weather and climate forecasting, a high-dimensional reservoir and a sufficient number of qubits on quantum hardware are required. The performance of quantum hardware is, however, limited by the presence of environmental and sampling noise. In this work, we study the effect of sampling noise on chaotic and turbulent systems, which exhibits extreme events. 
The findings of this paper are four-fold. First, we provide a mathematical overview of RF-QRC that involves a temporal memory with leaky integral, which also provides exponential smoothing of noisy states. This also makes the RF-QRC model scalable and suitable for temporal learning tasks. Second, we compare the effects of finite-sampling noise on quantum reservoir architectures with and without recurrence. We show that the framework of RF-QRC is more resilient to sampling noise than QRC with correlated noise. Third, we propose two methods based on SVD and signal filtering to suppress noise in reservoir activation signals. Our emulated results indicate that suppressing noise improves the training accuracy as highlighted by smaller mean-squared training errors and higher signal-to-noise (SNR) ratios. The methods of denoising applied in this work are general and the same analysis could be extended further by employing different advanced techniques for noise filtering to further improve the performance. Finally, we demonstrate our proposal by employing RF-QRC on multiple parallel QPUs on hardware backends, coupled with denoising techniques. The results indicate that denoising helps suppress finite sampling noise as well as other types of hardware noises. The proposed denoising can be extended to standard quantum reservoir computing architectures with recurrence and quantum extreme learning machines to suppress sampling and other types of noises for optimal training. This work opens up opportunities to employ quantum reservoir computing on quantum hardware for chaotic time-series forecasting. 

\begin{acknowledgments}
O.A. thanks Defne E. Ozan for fruitful discussions on classical reservoir computing. The authors acknowledge financial support from the UKRI New Horizon grant EP/X017249/1. L.M. is grateful for the support from the ERC Starting Grant PhyCo 949388, F.T. acknowledges support from the UKRI AI for Net Zero grant EP/Y005619/1.
\end{acknowledgments}

\begin{appendix}

\section{Three-dimensional Lorenz-63 model}\label{sec:AppL63}

One of the analysed system, Lorenz-63 \citep{75462} is a reduced order model of thermal convection flow. In this model, the fluid is heated uniformly from bottom and cooled from the top. Mathematically
\begin{equation}\label{eq:3A}
    \frac{dx_{1}}{dt} = \sigma \, (x_{2}-x_{1})
\end{equation}
\begin{equation}\label{eq:3B}
    \frac{dx_{2}}{dt} = x_{1} \, (\rho-x_{3}) - x_{2}, 
\end{equation}
\begin{equation}\label{eq:3C}
    \frac{dx_{3}}{dt} =  x_{1}x_{2}-\beta{x_{3}},
\end{equation}
where $\sigma$ , $\rho$ , $\beta$ are system parameters and we take [$\sigma$ , $\rho$ , $\beta$] = [10, 28, 8/3] to ensure chaotic behavior of the system. The largest Lyapunov exponent describes the non-linear dynamics \cite{racca_data-driven_2022}, which is $\Lambda=0.9$ for Lorenz-63 system and $1 LT = 1/0.9$ \citep{75462}. The time series data set is derived numerically via Runge-Kutta method and by taking a time-step size $dt=0.01$. The training time series comprises data points over a total time of 20 LT. For the training, the reservoir is evolved in an open loop to obtain reservoir states and calculate the $\pmb{W}_{out}$ matrix and associated training mean-squared error. 

\section{Nine-dimensional turbulent chaotic shear flow model}\label{sec:App_MFE}

To study turbulence, we consider a qualitative low-order model of turbulent shear flows, which is based on Fourier modes. Also known as the MFE model (`Moehlis, Faisst, and Eckhardt (MFE)' model), it is a non-linear model that captures the relaminarization and turbulent bursts \citep{moehlis_low-dimensional_2004}. Due to the non-linear nature of this model, the MFE model has been employed to study turbulence transitions and chaos predictability \citep{srinivasan_predictions_2019}. Mathematically, the MFE model can be described by the non-dimensional Navier-Stokes equations for forced incompressible flow
\begin{equation}\label{eq:19}
    \frac{{\partial}\pmb{v}}{{\partial} t} =  - (\pmb{v} \, . \, \pmb{\nabla} )\,\pmb{v} \,-\pmb{\nabla} \, {p}+ \frac{1}{Re} \Delta \pmb{v}  + \pmb{F}(y) , \quad \quad \pmb{\nabla}.\pmb{v} = 0
\end{equation}
where $\pmb{v} = (u, v, w)$ is the three-dimensional velocity vector, $p$ is the pressure, $Re$ is the Reynolds number, $\nabla$ is the gradient, and $\Delta$ is the Laplacian operator. $\pmb{F}(y)$ on the right-hand side is the sinusoidal body forcing term, which is $\pmb{F}(y) = \sqrt{2} \pi^{2}/(4 Re) \, \text{sin}(\pi y/2) \, \pmb{e_{x}} $. The body forcing term is applied between the plates along the $x,y$ direction of the shear. Furthermore, we consider a three-dimensional domain of size $L_{x}\times L_{y}\times L_{z}$ = $[4\pi,2,2\pi]$ and apply free slip boundary conditions at $y = L_{y}/2$, periodic boundary conditions at $x = [0 ; L_{x}]$ and $z = [0; L_{z}]$. The set of PDEs can be converted into ODEs by projecting the velocities onto Fourier modes as given by Eq.~\eqref{eq:20}
\begin{equation}\label{eq:20}
    \pmb{v}(\pmb{x} , t) = \sum_{i=1}^{9} \, a_{i}(t) \, \hat{\pmb{v}_{i}}\,(\pmb{x}).
\end{equation}
These nine decompositions for the amplitudes ${a_{i} (t)}$ are substituted into Eq.~\eqref{eq:19} to yield a set of nine ordinary differential equations as in \citep{moehlis_low-dimensional_2004}. The MFE system displays a chaotic transient, which in the long term converges to a stable laminar solution. We want to predict the turbulent burst of kinetic energy and chaotic transients, which are extreme events. We solve the MFE system ODEs using an RK4 solver with $dt = 0.25 $. The leading Lyapunov Exponent of the MFE model \citep{racca_data-driven_2022} is $\Lambda = 0.0163$. The length of each training time series is 65 LT. The resulting time series is used as input data $\pmb{u}_{in}(t)$ for training reservoir networks. 

\end{appendix}

\bibliography{references}% common bib file

\end{document}